\documentclass[aps,prl,10pt,twocolumn,superscriptaddress]{revtex4-1}

\usepackage[utf8]{inputenc} 
\usepackage[T1]{fontenc}
\usepackage{amsmath}
\usepackage{amssymb}
\usepackage{gensymb}
\usepackage{bbm}
\usepackage{graphicx}
\usepackage{framed}
\usepackage{color}
\usepackage{hyperref}
\usepackage{epstopdf}
\usepackage{dsfont}
\usepackage{amssymb}
\usepackage{amsmath}
\usepackage{amsthm}
\usepackage{wrapfig}
\usepackage{bm}
\usepackage{enumitem}

\DeclareUnicodeCharacter{00A0}{~}

\usepackage{pifont}

\newcommand{\ba}{\begin{eqnarray}}
\newcommand{\ea}{\end{eqnarray}}

\begin{document}

\title{Assessing data postprocessing for quantum estimation.}

\author{Ilaria Gianani}\email{ilaria.gianani@uniroma3.it}
\affiliation{Dipartimento di Scienze, Universit\`{a} degli Studi Roma Tre, Via della Vasca Navale 84, 00146, Rome, Italy}
\affiliation{Dipartimento di Fisica, Sapienza Universit\`{a} di Roma, P.le Aldo Moro, 5, 00185, Rome, Italy}

\author{Marco G. Genoni}
\affiliation{Dipartimento di Fisica, Università degli Studi di Milano, 20133, Milan, Italy}

\author{Marco Barbieri}
\affiliation{Dipartimento di Scienze, Universit\`{a} degli Studi Roma Tre, Via della Vasca Navale 84, 00146, Rome, Italy}
\affiliation{Istituto Nazionale di Ottica - CNR, Largo Enrico Fermi 6, 50125, Florence, Italy}

\begin{abstract}

Quantum sensors are among the most promising quantum technologies, allowing to attain the ultimate precision limit for parameter estimation. In order to achieve this, it is required to fully control and optimize what constitutes the hardware part of the sensors, i.e. the preparation of the probe states and the correct choice of the measurements to be performed. However careful considerations must be drawn also for the software components: a strategy must be employed to find a so-called optimal estimator. Here we review the most common approaches used to find the optimal estimator both with unlimited and limited resources. Furthermore, we present an attempt at a more complete characterization of the estimator by means of higher-order moments of the probability distribution, showing that most information is already conveyed by the standard bounds.

\end{abstract}

\maketitle

\section{Introduction}
The promises of quantum sensors \cite{GiovannettiNatPhot,degen17RMP} can be made possible on the one hand by the capabilities offered by novel physical devices and, on the other, by proper processing of the output signals. 
At a difference from celebrated examples of quantum computing, the output of a quantum sensor does not contain the sought answer in an explicit form. A further crucial step thus consists in choosing the optimal procedure to extract the parameters of interest from the measured quantities. Quantum metrology mostly focuses on the in-principle limitations associated to determine the more informative quantities and probe initialization. This does not exhaust the quest for optimal estimation, since adequate signal processing techniques have to be implemented. 
In operational terms, an estimator is needed, {\it i.e.} a way to establish a relation between the collected data and the values of the parameters. A key requirement for the estimator is to be unbiased, thus giving the "true" value of the parameters when taking the average over a large sample.

The problem at this point has long left the quantum realm, as it reduces to finding the best estimator of the parameters starting from a collection of ordinary data, i.e. the registered outcomes. Conversely, the quantum origin of the data still manifests in the benchmarking pertinent to the estimation. 
The Quantum Cram\'er-Rao Bound (QCRB) is a case in point \cite{MatteoIJQI}. The bound establishes a minimal variance for any unbiased estimator, only based on quantum formalism. This builds the connection between the quantum nature of the employed resources and the classical processing of the collected data. 
The closer the variance of the estimated parameter is to the QCRB, the higher the quality of the data analysis can be considered.

In this work we review examples of different strategies that have been followed to achieve optimal estimation in quantum photonics parameter estimation. In most cases, considerations extending well-beyond the merits on the specific implementation can be drawn. Also, we consider a more exhaustive characterization of the estimator which can be applied in quantum phase estimation beyond QCRB. To this end we inspect properties of probability distributions of phases in simulated experiments reaching beyond the mere inspection of the variance. 
We have adopted the bounds established by Barankin to higher order absolute moments \cite{barankin49}. Our finding is that the saturation of the QCRB conveys most of the information, thus confirming the exhaustiveness of this approach. 

\section{(Quantum) Parameter Estimation} \label{s:estimation}
We consider a classical statistical model described by a probability distribution $p(k|\lambda)$, where $\lambda$ is the parameter that we want to estimate, and $k$ denotes the outcome of the measurement that we have performed in order to deduce the value of $\lambda$. For simplicity in the following we will consider a discrete set of possible outcomes $k$. We now suppose that we repeat the measurement $M$ times, and thus we collect a set of measurement outcomes $\chi = \{ k_1, k_2, \dots, k_M \}$. In order to infer the parameter value $\lambda$ we have to build an estimator $\hat{\lambda}(\chi)$, {\it i.e.} a map associating possible sets of measurement outcomes $\chi$ to the possible values of the parameter $\lambda$. An estimator is defined as unbiased if satisfies the following property: $\mathbbm{E}[\hat{\chi}] := \sum_\chi p(\chi | \lambda) \hat{\lambda}(\chi) = \lambda$. Any unbiased estimator has to satisfy the following inequality, known as Cram\'er-Rao bound (CRB):
\begin{align}
\mathbbm{E}[(\hat{\lambda} - \lambda)^2] \geq \frac{1}{M \, \mathcal{F}[p(k|\lambda)]} \,,
\label{eq:CRB}
\end{align}
where 
\begin{align}
\mathcal{F}[p(k|\lambda)] = \sum_k p(k|\lambda) \left( \frac{\partial \log p(k|\lambda)}{\partial\lambda} \right)^2 \,,
\end{align}
is the so-called Fisher information. In the quantum realm, the classical statistical model is obtained from a quantum statistical model, described by a density operator $\varrho_\lambda$, via the Born rule $p(k|\lambda)= \hbox{Tr}[\varrho_\lambda \Pi_k]$, where the set of operators $\{\Pi_k\}$ defines a positive-operator valued measure (POVM). By optimizing over all the possible POVMs, that is over all the possible measurements allowed by quantum mechanics, one obtains the more fundamental quantum Cram\'er-Rao bound (QCRB) that depends on the quantum statistical model $\varrho_\lambda$ only \cite{MatteoIJQI,helstrom1976quantum,CavesBraunstein},
\begin{align}
\mathbbm{E}[(\hat{\lambda} - \lambda)^2] \geq \frac{1}{M \, \mathcal{F}[p(k|\lambda)]} \geq \frac{1}{M \, \mathcal{Q}[\varrho_\lambda]}
\label{eq:QCRB}
\end{align}
where $\mathcal{Q}[\varrho_\lambda] = \hbox{Tr}[\varrho_\lambda L_\lambda^2]$ is the quantum Fisher information, and the operator $L_\lambda$ denotes the symmetric logarithmic derivative (SLD) operator defined by the Lyapunov equation $2 \partial_\lambda \varrho_\lambda = L_\lambda \varrho_\lambda + \varrho_\lambda L_\lambda$.
Both inequalities in Eq. (\ref{eq:QCRB}) are in fact achievable in the single-parameter scenario: while the QCRB can be saturated by choosing the optimal measurement, corresponding to the eigenbasis of the SLD operator $L_\lambda$, the (classical) CRB can be saturated by choosing an {\it optimal} estimators. In the next sections we will first review some widely used estimators, along with some applications in quantum experiments, and discussing their optimality in terms of the CRB. We will then discuss a generalization of the CRB for higher order moments of the estimator, and we will discuss the performance of Bayesian estimation in terms of these generalized bounds.
\section{Bayesian Estimation} \label{s:bayes}
We consider the classical statistical model $p(k|\lambda)$ and the data sample $\chi = \{k_1,k_2, \dots, k_M\}$. We can evaluate the likelihood function of the sample as
$$p(\chi | \lambda) = \prod_{j=1}^M p(k_j | \lambda)\,.$$ 
By means of Bayes theorem, we can then obtain the {\it a-posteriori} probability
\begin{align}
P(\lambda | \chi) = \frac{p(\chi |\lambda) p(\lambda) }{p(\chi)} \,,
\end{align}
where $p(\lambda)$ is the {\it a-priori} probability distribution; this can in general assumed flat, if no a-priori information is available on the parameter $\lambda$, while $p(\chi)$ is simply a normalization constant. The Bayesian estimator is the mean of the a-posteriori probability, i.e.
\begin{align}
\hat{\lambda}_{B}(\chi)  = \mathbbm{E}[\lambda] := \int d\lambda \, \lambda \, P(\lambda |\chi) \,.
\end{align}
It can be shown that in the asymptotic limit of large number of measurements ($M \to \infty$), the a-posteriori distribution $p(\lambda|\chi)$ tends to a Gaussian distribution \cite{braunstein92}, and that the Bayesian estimator is both unbiased and optimal: the  variance of the Gaussian distribution, defined as 
$$
\hbox{Var}(\lambda) = \int d\lambda \, \lambda^2 \, P(\lambda |\chi) - \left(\hat{\lambda}_{B}(\chi) \right)^2 \,,
$$
becomes equal to $1/(M \mathcal{F}[p(k|\lambda)])$, and thus saturates the CRB. 

\begin{figure}[t]
\centering
\includegraphics[width=1\columnwidth]{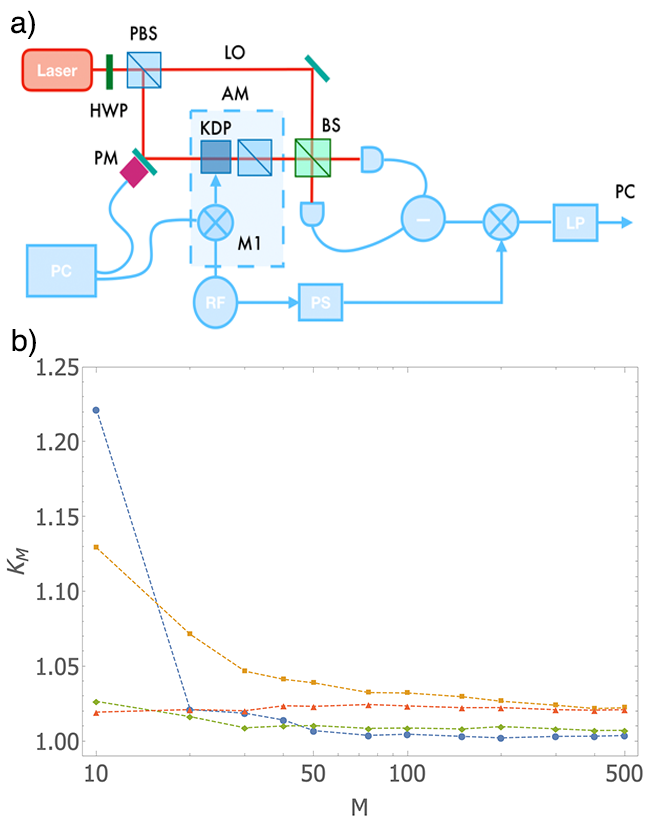}
\caption{Bayesian estimation in the presence of large phase diffusion. a) Experimental Setup.b) Variances of the estimator normalized by the CRB, as a function of the number of events M and for different values of the mean number of photons N and the noise parameter $\Delta$. Blue : N = 0.90, $\Delta = \pi/18$ rad; yellow: N = 0.90, $\Delta = \pi/9$ rad; green: N = 4.12,$\Delta = \pi/18$ rad; red:N = 4.12,$\Delta = \pi/9$ rad.}
\label{geno}
\end{figure}

This technique has been widely discussed both theoretically and experimentally in the quantum regime (see for example  \cite{hradil95,hradil96,olivares09,brivio10,genoni12,blandino12}).
In particular in Refs. \cite{brivio10,genoni12}, it has been employed by M.G.A. Paris's group to investigate phase estimation in the presence of phase diffusion both for discrete variable and continuous variable systems. These represent one the primary examples of parameter estimation affected by noise. As regards the continuous-variable case in \cite{genoni12}, probe light is initialized in the state $\rho_0$, in order to estimate a small optical phase $\phi\simeq 0$; fluctuations of the latter result in different runs sensing distinct values of $\phi$, assumed to be distributed Gaussianly with zero mean and known width $\sqrt{2}\Delta$, characterizing the phase diffusion rate. 
The experiment has made use of a proof-of-principle setup, in which the phase to be estimated was inserted between two arms of a Mach-Zehnder interferometer, and phase diffusion was applied in a controlled manner by means of a mirror mounted on a piezoelectric actuator (Fig. \ref{geno} a). The input probe states were coherent states, as they perform close to optimal, especially under strong-noise condition. The measurement, chosen to optimize the QCRB, is homodyne detection set to measure the amplitude quadrature of the probe. 

The data follows closely the probability distribution predicted by the model, hence Bayesian estimation can be reliably applied to extract the value of the phase, as well as the variance. This permits to evaluate directly the performance of the estimator, without being affected by artifacts and bias due to uninformed modelling. The analysis reveals that saturation of the classical CRB is achieved with $M\geq100$ repetitions (Fig. \ref{geno} b), for different values of noise and probe intensity, a value that poses no experimental challenges. This is evidence for the effectiveness of Bayesian estimation.

\subsection{Bayesian estimation in the presence of noise} \label{s:optica}

The results presented in Ref \cite{genoni12} indicate that accurate modelling is required to benefit from the strengths of Bayesian estimation. When dealing with realistic devices, the outcome probabilities are governed by multiple parameters: while not all of them will be informative on the physical process, they all need to be assessed correctly to avoid biases in the estimation.
Nevertheless, the Bayesian approach can be used to provide unbiased estimation by monitoring together with the sought parameter, further ones conveying the information on probe and measurement employed for the estimation. This allows to reduce the effect of biases resulting from a scarce control on the probe quality. This is the case in point in \cite{roccia18}. \\

Multiparameter estimation has general scope beyond this application, and many instances have shown a trade-off in the achievable precision on the parameters to be estimated ,including multiple phases, phase and phase diffusion, etc. \cite{ mihai, nico19, animesh}. Such trade-offs do not impact on the capabilities of the Bayesian technique to provide a reliable estimator for many parameters at once. 

\begin{figure}[t]
\centering
\includegraphics[width=1\columnwidth]{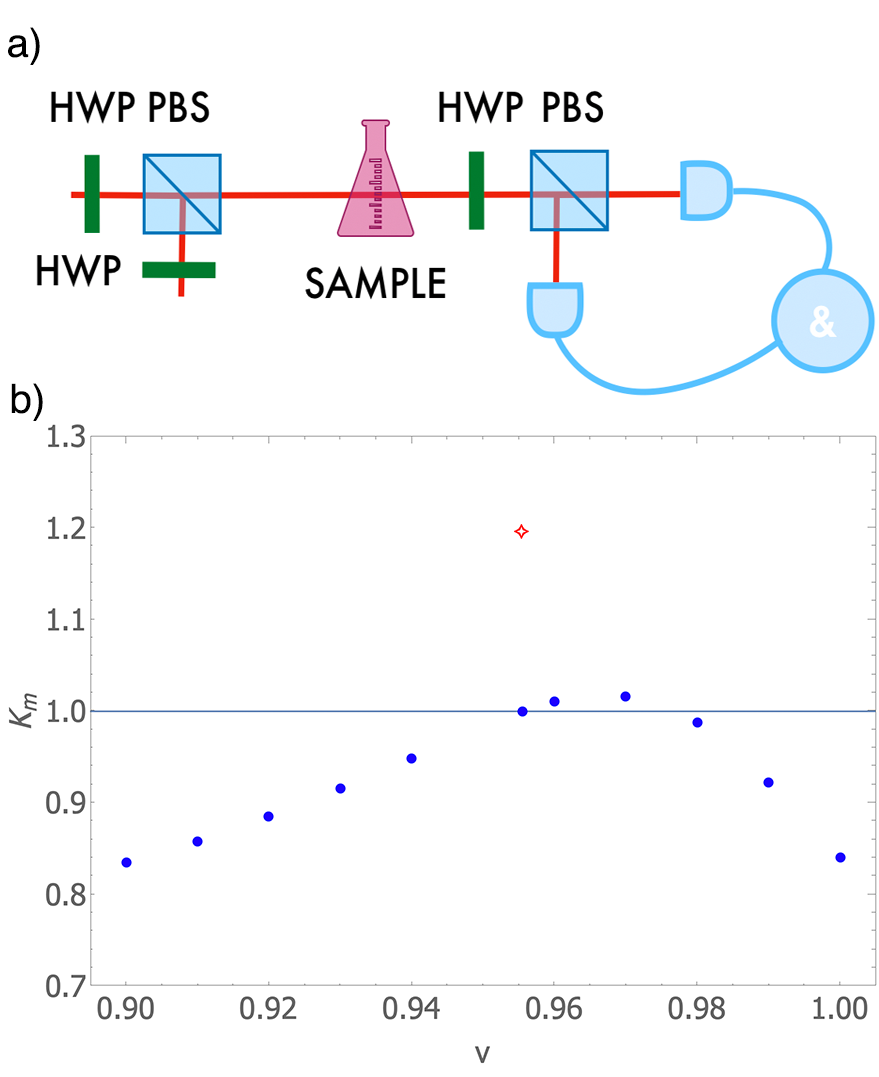}
\caption{Bayesian estimation with limited visibility. a) Experimental setup. b) Variances of the estimator $\hat{\phi}$ normalized by the CRB. The blue dots are obtained by single parameter Bayesian estimation with fixed visibility, while the red diamond is obtained using the multiparameter approach. 
Note that the latter figure differs from the one originally in the supplementary information of \cite{roccia18} due to a more reliable definition of the limits of integration. }
\label{optica}
\end{figure}

Ref \cite{roccia18} has employed N00N states with limited visibility to measure the optical activity of sugars solutions in water. The multiparameter approach addressed the simultaneous measurement of an optical phase related to the activity, and the visibility. This is shown to be a preferable option with respect to state pre-calibration, as this may change in different experimental conditions. The setup in Fig. \ref{optica} a) describes the sensing apparatus: a N00N state in the circular polarization basis is sent through a polarization interferometer. A sample containing the solution to be measured, placed in the optical path, introduces a birefringent phase shift as a result of the optical activity of the specimen.  
Coincidence counts are then recorded for different projection measurements, obtaining the post-selected probabilities: 
\begin{align}
p(\theta\vert 1; \phi; v)= \frac{1}{4}\left(1+ v \,\cos(2\phi-8\theta)\right) \,,
\label{eq:proboptica}
\end{align}
where $\phi$ is the phase to be estimated, $v$ is the visibility which determines the quality of the probe, and $\theta$= $0, \pi/16, \pi/8, 3\pi/16$, are used to obtain the correct normalization. 

The estimation is performed for a solution of sucrose and for one of fructorse, the former exhibiting a dexorotatory optical activity, the latter a levorotatory one.  The collected data are thus processed by extending the Bayesian approach to account for two parameters.

Further, the data for the sucrose solution are evaluated with a Bayesian single parameter routine, implemented by fixing an arbitrary value for the visibility $v$. By comparing the two estimation against the CRB, it is possible to quantify the bias introduced by a wrong assumption on the visibility. Fig. \ref{optica} b) shows the comparison of the variances, normalizes by the CRB, for different values of $v$ imposed in the single parameter analysis, together with the outcome of the multiparameter estimation.

\section{Maximum likelihood}

A different approach to attain an optimal estimator relies on the use of a maximum likelihood function \cite{paris01}. This quantifies how likely a value of the parameter can describe the collected data. This approach is clearly illustrated taking as a guiding example the experiment in Ref.\cite{john18}.
There, C. Silberhorn's group experimentally tackles the optimal estimation of spectral and temporal separation of two mutually-incoherent single-photon sources. 
\begin{figure}[b!]
\centering
\includegraphics[width=1\columnwidth]{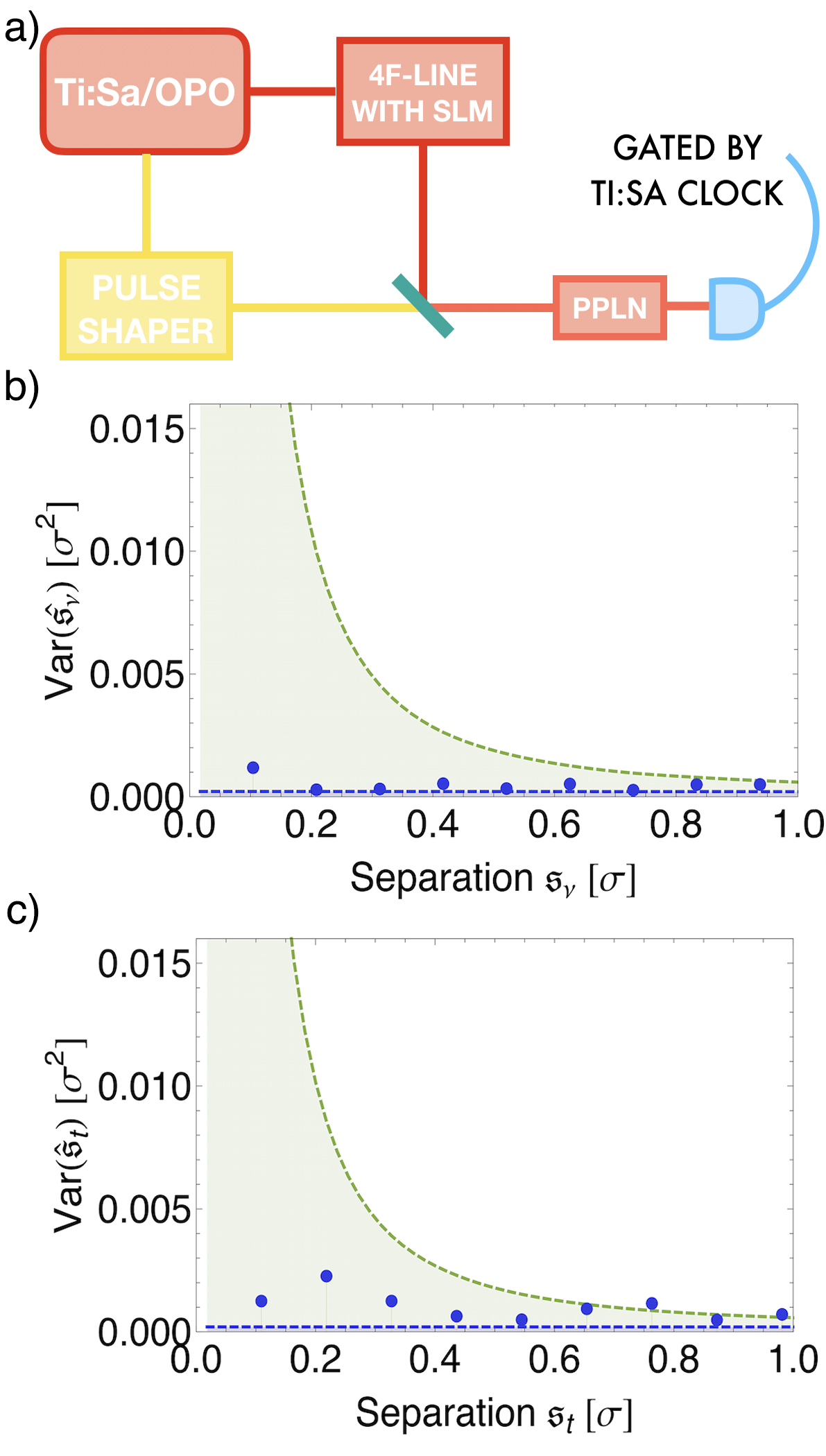}
\caption{Parameter estimation through time-frequency mode-selecting measurements. a) Experimental setup. The 1540 nm output of an OPO (yellow) is sent through a pulse shaper to create the desired separation $\mathfrak{s}$, while the 875 nm output (red) is sent through a 4f-line and shaped into HG modes by means of a spatial light modulator (SLM). The upconversion is performed with a PPLN waveguide, and the photons are then detected with an avalanche phodtodiod gated by the clock pulse from the Ti:Sa. b) and c) Mean square error of the estimator $\hat{\mathfrak{s}}$ for the frequency (time) separation, performed with M=20 000 events. The green dashed curve is the CRB for the intensity measurement, while the blue dashed curve is the QCRB. The latter two figures have been adapted from \cite{john18}.}
\label{xstine}
\end{figure}

The manuscript considers two Gaussian pulses, each with root-mean square (rms) width $\sigma_\nu$, spectrally separated by an amount $\mathfrak{s}_\nu$ or, alternatively, separated in time by $\mathfrak{s}_t$. While the QCRB is constant \cite{qcrbxstine} with the value of the separation, performing spectral intensity measurements yields a CRB which diverges as the ratio $\mathfrak{s}_\nu$/$\sigma_\nu$ goes to zero. 
Approaching the QCRB for every $\mathfrak{s}_\nu$/$\sigma_\nu$ becomes possible when performing measurements on a set of projections with definite parity. For Gaussian pulses the optimal basis is identified with Hermite-Gaussian (HG) modes \cite{luisxstine2}, which can be further restricted to the first two modes for all practical purposes, if $\mathfrak{s}_\nu$/$\sigma_\nu$ $< 1$. Higher order modes must be considered for $\mathfrak{s}_\nu$/$\sigma_\nu$ $> 1$ \cite{luisxstine}, but this sophisticated strategy yields no advantage with respect the standard intensity measurement in such regime. Analogous conclusions can be drawn in the time domain. 

In order to project on HG states, the quantum pulse gate \cite{eckstein11} described in Fig. \ref{xstine} a) is implemented. 
This consist in selectively upconverting the Gaussian pulses with a strong pump shaped in HG modes in a long nonlinear waveguide. The resulting upconverted signal corresponds to the projection of the Gaussian beams onto the selected HG mode of the pump. 

The experiment is performed by projecting on the first three HG modes, namely $HG_0, HG_1$, and $HG_2$.  In order to obtain an unbiased estimator towards the imperfection of the setup, a calibration is performed via quantum tomography with known separations. The three HG modes are considered as a complete basis for the measured probability $p_j$ (with j= 0,1,2) for each experimental projection.

The likelihood function manipulates the experimental probabilities, which depend explicitly on the sought parameter, in order to identify the most probable outcome as the estimator. In the case of Ref.\cite{john18}, the outcomes are in the form of a triplet $\{n_0,n_1,n_2\}$ of counts corresponding to the three projections. By assuming that these follow a Poissonian statistics, we can use the same form of the likelihood function as in Bayesian estimation 
\begin{equation}
    \mathcal{L}(\mathfrak{s})=\prod_{j=0}^{2}p_j(\mathfrak{s})^{n_j},
\end{equation}
or, equivalently, its logarithm. The estimator is then the value $\hat{\mathfrak{s}}$ that maximizes $\mathcal{L}(\mathfrak{s})$. 
Differently from Bayesian estimation, we get a single value of $\hat{\mathfrak{s}}$ from a single measurement: the uncertainty then needs to be evaluated by repeating the measurement multiple times, and assessing the statistics of the individual estimates. In Ref.\cite{john18}, the variance of $\hat{\mathfrak{s}}$ is evaluated from 60 repetitions.  

Fig. \ref{xstine} b) - c) show the mean square errors, compared with the CRB for intensity measurements and with the QCRB, both for spectrally and temporally separated pulses. The maximum likelihood approach provides an adequate estimator, relying, as in the previous case, on the knowledge of the actual detection probabilities.

\section{Machine Learning}
Bayesian and maximum likelihood strategies demonstrate their effectiveness in the limit of a large number of repetitions $M$, although this regime can be achieved in the experiment. However these techniques do not work well when the resources are scarce \cite{rubio19}. Here we do not refer to the fact that the expected variance increases with $1/M$,  but to the inability of finding a proper estimator.

For this purpose, machine learning represents a precious tool; applications to optical phase estimation have been thoroughly investigated by the group of F. Sciarrino~\cite{lumino18}. In their work, they compare machine-learning assisted protocols to adaptive phase estimation with a limited number of repetitions $M$. Their experimental setup consists in a Mach-Zehnder interferometer with an unknown phase $\phi$ on one arm, and an adaptive phase $\Phi$ on the other. The benchmark protocol is the particle guess heuristic (PGH) approach~\cite{wiebe16}, consisting in updating the feedback phase at each step with a random guess selected from the posterior distribution.

The first algorithm investigated is the particle swarm optimization (PSO). This was originally introduced in \cite{pso1,pso2}, and consists in updating at each step the feedback phase by the rule: $\Phi_k = \Phi_{k-1}-(-1)^{x_{k-1}}\Delta\Phi_k$, where $x_{k-1}=0,1$ is the outcome of the k-1 measurement. The $M$-size vector of the  $\Delta\Phi_k$, called {\it policy} is evaluated in advance by mapping the policies into the evolution of $n$ particles. The optimality criterion is the minimization of the Holevo variance, a quantity related to the QCRB. The approach is effective, but, since its complexity scales as $O(N^6)$, and it becomes uneffective for  $ M \geq  45-50$. 

The second method considered is Gaussian optimal (GO) Bayesian estimation. It is based on the assumption that the prior distribution $P(\phi)$ is Gaussian with mean $\mu$ and variance $\sigma^2$, a condition which is typically satisfied from  $N\simeq 10$. In this case, it is indeed possible to provide an analytic expression (that holds for $\sigma \geq 0.921$) for updating the feedback phase at each step in order to minimize the variance of the a-posteriori phase distribution. The algorithm is demonstrated to yield the optimal and unbiased estimator for every phase in the $[0, 2\pi]$ interval. The GO algorithm is less demanding than the PSO, as the computational resources scale with $O(N)$. 

Both machine-learning methods demonstrate in the experiment performance that are superior to the simple PHG strategy, even in the presence of noise~\cite{lumino18}.

\section{Beyond the CRB: Achievability of generalized CRBs in a quantum phase-estimation protocol}

The only criterion which is generally employed for assessing optimality of an estimator is whether its variance reaches the QCRB. This overlooks other information that may be present in the complete parameter distribution, and that can be captured by looking at higher-order moments. Following the same procedure as in Sec. \ref{s:estimation}, we consider a classical statistical model described by conditional probabilities $p(k|\lambda)$ of observing the outcome $k$ given the value of a parameter to be estimated $\lambda$, and a sample of observed data $\chi=\{k_1,k_2,\dots,k_M\}$. In \cite{barankin49} Barankin has derived a family of generalized bounds for the $\beta$-th central absolute moments of any estimator $\Sigma_\beta = \mathbbm{E}\left[| \hat\lambda (\chi) - \lambda |^\beta \right]$
\begin{align}
&\Sigma_\beta
\geq \frac{1}{M^\frac{\beta}{2}  \, \mathcal{F}_\alpha [ p(k |\lambda)]^{\frac{\beta}{\alpha}} }\,,\\
\label{eq:genCRB}
\nonumber
&\textrm{with} \,\,\, \alpha,\beta>1 \,\,\,\,\,\textrm{and}\,\,\,\, \frac{1}{\alpha} + \frac 1\beta = 1 \,,
\nonumber
\end{align}
where we have defined a family of generalized Fisher information functions
\begin{align}
\mathcal{F}_\alpha [ p(k|\lambda) ] = \sum_k p(k|\lambda) \, \left| \frac{\partial \log p(k|\lambda)}{\partial\lambda} \right|^\alpha \,.
\end{align}
By choosing $\alpha=\beta=2$ the inequality above reduces to the (classical) CRB in Eq. (\ref{eq:CRB}), that poses the ultimate bound on the variance of any unbiased estimator. In general, if we fix the moment's order $\beta$ and thus the corresponding Fisher information order $\alpha= \beta/(\beta-1)$, we can define the quantity
\begin{align}
\Xi_\beta := \Sigma_\beta \, M^{\frac{\beta}{2}} \, \mathcal{F}_\alpha [ p(k |\lambda)]^{\frac\beta\alpha} \geq 1
\label{eq:barankinXi}
\end{align}
As we pointed out in Sec. \ref{s:bayes}, Bayesian estimator is asymptotically optimal in terms of the Cram\'er-Rao bound, i.e. for a large number of measurements $M \gg 1$, one obtains that $\Xi_2 \approx 1$, and the a-posteriori Bayesian distribution tend to a Gaussian with a variance that is consequently equal to the (rescaled) inverse Fisher information. %

%

\begin{figure}[t]
\includegraphics[width=0.9\columnwidth]{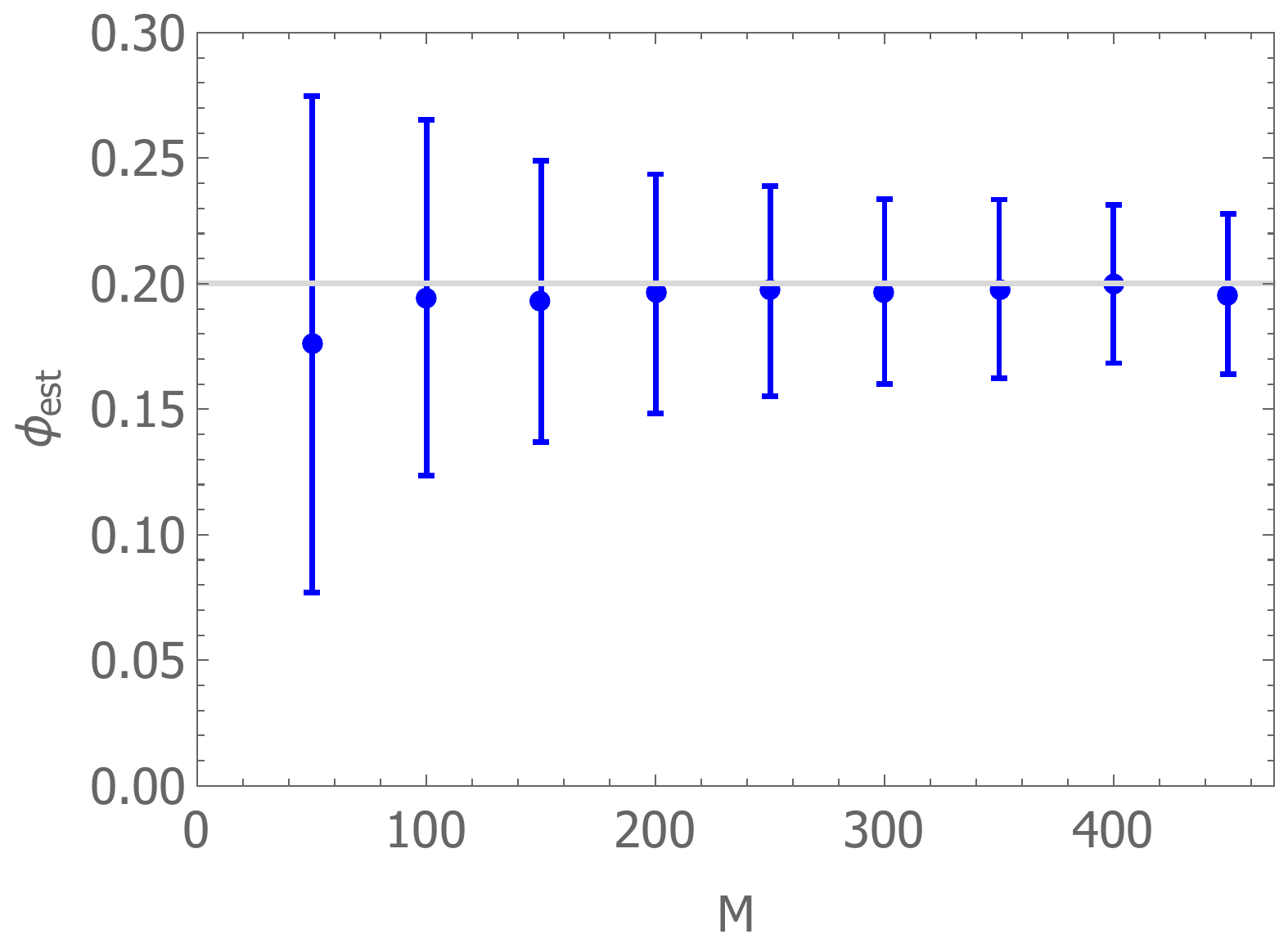} 
\caption{Average estimated value $\hat{\phi}_B$ via Bayesian estimation as a function of the number of measurements $M$, obtained by averaging over $N=500$ simulated experiments. The light-grey line represents the true value of the phase to be estimated $\phi=0.2$ (visibility is fixed to $v=0.9$).
}
\label{f:PhiEst}
\end{figure}
Here we use these generalized bounds in Eq. (\ref{eq:barankinXi}) for higher orders moments to assess the quantum phase-estimation protocol with limited visibility experiment we have reviewed in Sec. \ref{s:optica}, and we will discuss their attainability. The measurement outcome probabilities, defining the classical statistical model, in Eq. (\ref{eq:proboptica}) can be recasted as 
\begin{align}
p_v(k | \phi ) = \frac{1}{4} \left[ 1 + v \, \cos(2\phi - k \pi /2) \right] \,\,\,\,\, k=\{0,1,2,3\}.
\end{align}

\begin{figure}[]
\begin{center}

\includegraphics[width=0.8\columnwidth]{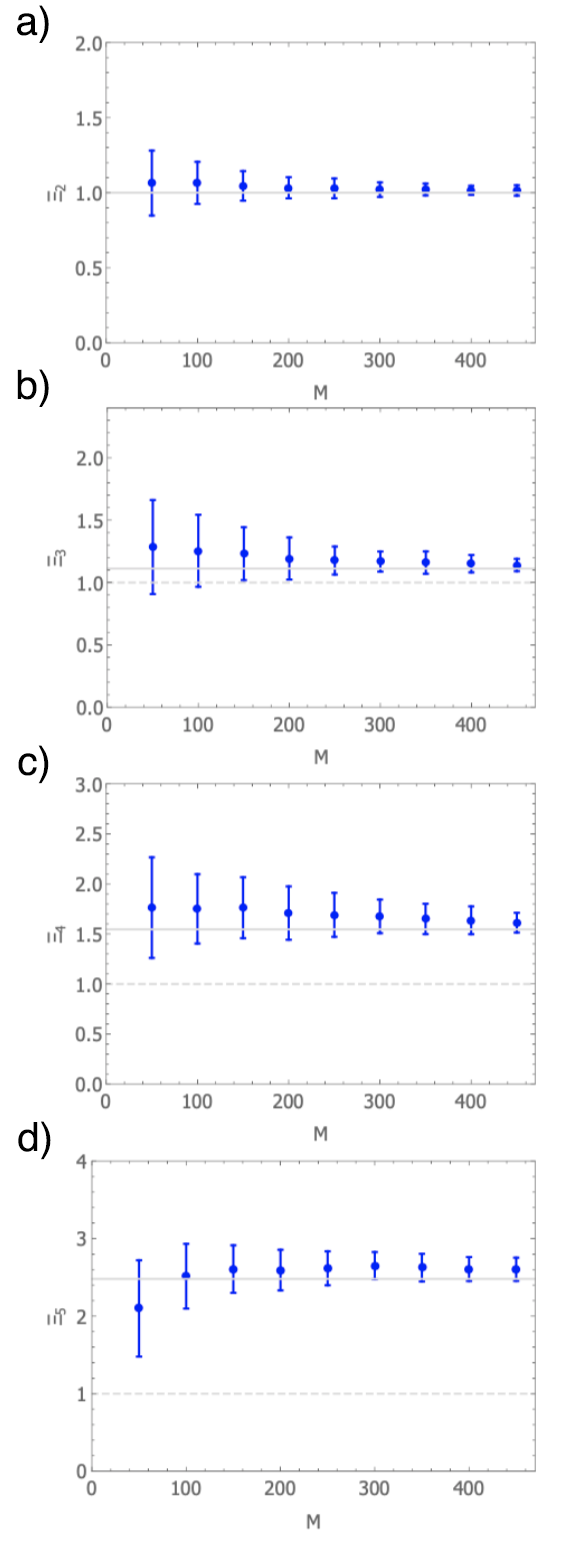} 
    
\end{center}
\caption{Average value of the quantity $\Xi_\beta$ corresponding to absolute central moments of the Bayesian a-posteriori distribution as a function of the number of measurements $M$, obtained by averaging over $N=500$ experiments. The light-grey solid line represents the expected limiting value for Gaussian distributions in Eq. (\ref{eq:XiGaussian}), while the light-grey dashed line represents the ultimate limit. Experimental parameters are set as follows: $\phi=0.2$, $v=0.9$. From top to bottom: $\beta=\{2,3,4,5\}$.}
\label{f:Xi}
\end{figure}

By means of these probabilities, and given a data sample $\chi$, we can construct the Bayesian a-posteriori probability $P(\phi | \chi)$ and calculate the central absolute moments of different order $\beta$. More in detail, we have simulated experiments involving different number of measurements, up to $M=450$ measurements, and in each experiment we have evaluated the quantity $\Xi_\beta$ for $\beta = \{2,3,4,5\}$. In Fig. \ref{f:PhiEst} we have first plotted the average estimated value of the unknown phase, showing how increasing the number of measurements it approaches the true value: the estimator is not biased.

In Fig.~\ref{f:Xi} a) we have reported the saturation of the standard CRB, corresponding to the value of $\beta=2$. The simulated parameters are: visibility $v=0.9$, true value of the phase $\phi=0.2$, and each experiment is simulated 500 times; the reported values are the average over these simulations and we have also reported the associated standard deviation. As expected, performing experiment with around 200 runs ensures the saturation of the CRB. This figure is of the same order as the one in Fig \ref{geno}.

In Fig.~\ref{f:Xi} b) the results for $\beta = 3$ are shown. We notice that the saturation occurs on a similar scale as for the variance, however the limit value does not correspond to the ultimate bound. Indeed, we find that it converges to the expected moment of a Gaussian distribution whose width is at the CRB. 
The bound on this moment is thus not as informative as for the variance, but looking at the Gaussianity still yields some insight.  

Fig.~\ref{f:Xi} c) and d) show the results for $\beta= 4,5 $ respectively. We can draw the same qualitative conclusions as for the previous case, while noting a slower convergence to the values of a Gaussian distribution. The departure from the ultimate bounds is more pronounced as the order increases. 

These results can be rigorously proved by analyzing the a-posteriori Bayesian distribution $P(\phi | \chi)$ that, as we pointed out before, for large number of measurements $M$ tends to a Gaussian distribution. One can indeed check that the known relationships for the central absolute moments of Gaussian distributions
\begin{align}
\Sigma_\beta = (\Sigma_2)^{\frac\beta2} \, (\beta -1)!!  \, \cdot \left\{
                \begin{array}{lc}
                  \sqrt{\frac{2}{\pi}} & \textrm{if}\,\,\beta\,\,\textrm{is odd} \\                  
		    1 & \textrm{if}\,\,\beta\,\,\textrm{is even}
                \end{array}
              \right. \label{eq:normalmoments}
\end{align}
are approximately satisfied by the a-posteriori distributions, as we increase the number of measurements $M$.  
As a consequence, we find that the quantities $\Xi_\beta$ is obliged to approach the following limiting values, 
\begin{align}
\Xi_\beta  &\stackrel{M \gg 1}{\approx} 
\left(\frac{\mathcal{F}_\alpha^{\beta/\alpha}}{\mathcal{F}_2^{\beta/2}}\right) (\beta -1)!!  \, \cdot \left\{
                \begin{array}{lc}
                  \sqrt{\frac{2}{\pi}} & \textrm{if}\,\,\beta\,\,\textrm{is odd} \\                  
		    1 & \textrm{if}\,\,\beta\,\,\textrm{is even}
                \end{array}
              \right.
\label{eq:XiGaussian}
\end{align}
where, as usual $\alpha=\beta/(\beta-1)$ and we used the relationship in Eq. (\ref{eq:normalmoments}) and the fact that the variance converges as $\Sigma_2 \approx 1/(M \mathcal{F}_2)$. This validates the results plotted in Fig. \ref{f:Xi} where we have highlighted the tighter limiting values larger than one, that are approached by $\Xi_\beta$ as the number of measurements $M$ is increased.
\section{Conclusion}
Post-processing of measurement data is a crucial step in parameter estimation. In particular it becomes fundamental to attain the promised ultimate precision achievable in quantum sensing protocols, after the optimal probe state and the optimal measurement strategy have been carefully chosen. Here we have introduced and discussed, by means of various examples, two of the most common strategies used in the asymptotic regime, i.e. Bayesian estimation and the maximum likelihood approach. Furthermore, we have presented more involved techniques based on machine learning that have to be implemented when the resources (i.e. the size of the data sample) are limited. 
Finally, we have explored the generalized CRBs to asses higher-order moments of the Bayesian a-posteriori distribution, showing that they do not result to be as informative as the (standard) CRB, and the predominant effect results from the Gaussianity of the a-posteriori distribution.
\section*{Acknowledgements}

The authors would like to thank M.G.A. Paris and V. Ansari for providing the material for Fig. \ref{geno} and \ref{xstine}, and J. M. Donohue, F. Albarelli, N. Spagnolo, M. Sbroscia and V. Cimini for fruitful discussions. 

\bibliographystyle{IEEEtran}
\bibliography{barankin.bib}

\begin{thebibliography}{10}
\providecommand{\url}[1]{#1}
\csname url@samestyle\endcsname
\providecommand{\newblock}{\relax}
\providecommand{\bibinfo}[2]{#2}
\providecommand{\BIBentrySTDinterwordspacing}{\spaceskip=0pt\relax}
\providecommand{\BIBentryALTinterwordstretchfactor}{4}
\providecommand{\BIBentryALTinterwordspacing}{\spaceskip=\fontdimen2\font plus
\BIBentryALTinterwordstretchfactor\fontdimen3\font minus
  \fontdimen4\font\relax}
\providecommand{\BIBforeignlanguage}[2]{{%
\expandafter\ifx\csname l@#1\endcsname\relax
\typeout{** WARNING: IEEEtran.bst: No hyphenation pattern has been}%
\typeout{** loaded for the language `#1'. Using the pattern for}%
\typeout{** the default language instead.}%
\else
\language=\csname l@#1\endcsname
\fi
#2}}
\providecommand{\BIBdecl}{\relax}
\BIBdecl

\bibitem{GiovannettiNatPhot}
\BIBentryALTinterwordspacing
V.~Giovannetti, S.~Lloyd, and L.~Maccone, ``{Advances in quantum metrology},''
  \emph{Nat. Photonics}, vol.~5, no.~4, p. 222, 2011. [Online]. Available:
  \url{http://www.nature.com/doifinder/10.1038/nphoton.2011.35}
\BIBentrySTDinterwordspacing

\bibitem{degen17RMP}
\BIBentryALTinterwordspacing
C.~L. Degen, F.~Reinhard, and P.~Cappellaro, ``Quantum sensing,'' \emph{Rev.
  Mod. Phys.}, vol.~89, p. 035002, Jul 2017. [Online]. Available:
  \url{https://link.aps.org/doi/10.1103/RevModPhys.89.035002}
\BIBentrySTDinterwordspacing

\bibitem{MatteoIJQI}
\BIBentryALTinterwordspacing
M.~G.~A. Paris, ``{Quantum Estimation for Quantum Technology},'' \emph{Int. J.
  Quant. Inf.}, vol.~07, no. supp01, p. 125, jan 2009. [Online]. Available:
  \url{http://www.worldscientific.com/doi/abs/10.1142/S0219749909004839}
\BIBentrySTDinterwordspacing

\bibitem{barankin49}
\BIBentryALTinterwordspacing
E.~W. Barankin, ``Locally best unbiased estimates,'' \emph{The Annals of
  Mathematical Statistics}, vol.~20, no.~4, pp. 477--501, 1949. [Online].
  Available: \url{http://www.jstor.org/stable/2236306}
\BIBentrySTDinterwordspacing

\bibitem{helstrom1976quantum}
C.~W. Helstrom, \emph{{Quantum Detection and Estimation Theory}}.\hskip 1em
  plus 0.5em minus 0.4em\relax New York: Academic Press, 1976.

\bibitem{CavesBraunstein}
\BIBentryALTinterwordspacing
S.~L. Braunstein and C.~M. Caves, ``{Statistical distance and the geometry of
  quantum states},'' \emph{Phys. Rev. Lett.}, vol.~72, no.~22, p. 3439, may
  1994. [Online]. Available:
  \url{https://link.aps.org/doi/10.1103/PhysRevLett.72.3439}
\BIBentrySTDinterwordspacing

\bibitem{braunstein92}
\BIBentryALTinterwordspacing
S.~L. Braunstein, ``How large a sample is needed for the maximum likelihood
  estimator to be approximately gaussian?'' \emph{Journal of Physics A:
  Mathematical and General}, vol.~25, no.~13, pp. 3813--3826, jul 1992.
  [Online]. Available:
  \url{https://doi.org/10.1088%2F0305-4470%2F25%2F13%2F027}
\BIBentrySTDinterwordspacing

\bibitem{hradil95}
\BIBentryALTinterwordspacing
Z.~c.~v. Hradil, ``Estimation of counted quantum phase,'' \emph{Phys. Rev. A},
  vol.~51, pp. 1870--1873, Mar 1995. [Online]. Available:
  \url{https://link.aps.org/doi/10.1103/PhysRevA.51.1870}
\BIBentrySTDinterwordspacing

\bibitem{hradil96}
\BIBentryALTinterwordspacing
Z.~Hradil, R.~My\ifmmode~\check{s}\else \v{s}\fi{}ka,
  J.~Pe\ifmmode~\check{r}\else \v{r}\fi{}ina, M.~Zawisky, Y.~Hasegawa, and
  H.~Rauch, ``Quantum phase in interferometry,'' \emph{Phys. Rev. Lett.},
  vol.~76, pp. 4295--4298, Jun 1996. [Online]. Available:
  \url{https://link.aps.org/doi/10.1103/PhysRevLett.76.4295}
\BIBentrySTDinterwordspacing

\bibitem{olivares09}
\BIBentryALTinterwordspacing
S.~Olivares and M.~G.~A. Paris, ``Bayesian estimation in homodyne
  interferometry,'' \emph{Journal of Physics B: Atomic, Molecular and Optical
  Physics}, vol.~42, no.~5, p. 055506, feb 2009. [Online]. Available:
  \url{https://doi.org/10.1088%2F0953-4075%2F42%2F5%2F055506}
\BIBentrySTDinterwordspacing

\bibitem{brivio10}
\BIBentryALTinterwordspacing
D.~Brivio, S.~Cialdi, S.~Vezzoli, B.~T. Gebrehiwot, M.~G. Genoni, S.~Olivares,
  and M.~G.~A. Paris, ``Experimental estimation of one-parameter qubit gates in
  the presence of phase diffusion,'' \emph{Phys. Rev. A}, vol.~81, p. 012305,
  Jan 2010. [Online]. Available:
  \url{https://link.aps.org/doi/10.1103/PhysRevA.81.012305}
\BIBentrySTDinterwordspacing

\bibitem{genoni12}
\BIBentryALTinterwordspacing
M.~G. Genoni, S.~Olivares, D.~Brivio, S.~Cialdi, D.~Cipriani, A.~Santamato,
  S.~Vezzoli, and M.~G.~A. Paris, ``Optical interferometry in the presence of
  large phase diffusion,'' \emph{Phys. Rev. A}, vol.~85, p. 043817, Apr 2012.
  [Online]. Available:
  \url{https://link.aps.org/doi/10.1103/PhysRevA.85.043817}
\BIBentrySTDinterwordspacing

\bibitem{blandino12}
\BIBentryALTinterwordspacing
R.~Blandino, M.~G. Genoni, J.~Etesse, M.~Barbieri, M.~G.~A. Paris, P.~Grangier,
  and R.~Tualle-Brouri, ``Homodyne estimation of gaussian quantum discord,''
  \emph{Phys. Rev. Lett.}, vol. 109, p. 180402, Nov 2012. [Online]. Available:
  \url{https://link.aps.org/doi/10.1103/PhysRevLett.109.180402}
\BIBentrySTDinterwordspacing

\bibitem{roccia18}
\BIBentryALTinterwordspacing
E.~Roccia, V.~Cimini, M.~Sbroscia, I.~Gianani, L.~Ruggiero, L.~Mancino, M.~G.
  Genoni, M.~A. Ricci, and M.~Barbieri, ``Multiparameter approach to quantum
  phase estimation with limited visibility,'' \emph{Optica}, vol.~5, no.~10,
  pp. 1171--1176, Oct 2018. [Online]. Available:
  \url{http://www.osapublishing.org/optica/abstract.cfm?URI=optica-5-10-1171}
\BIBentrySTDinterwordspacing

\bibitem{mihai}
\BIBentryALTinterwordspacing
M.~D. Vidrighin, G.~Donati, M.~G. Genoni, X.-M. Jin, W.~S. Kolthammer, M.~S.
  Kim, A.~Datta, M.~Barbieri, and I.~A. Walmsley, ``Joint estimation of phase
  and phase diffusion for quantum metrology,'' \emph{Nature Communications},
  vol.~5, pp. 3532 EP --, 04 2014. [Online]. Available:
  \url{https://doi.org/10.1038/ncomms4532}
\BIBentrySTDinterwordspacing

\bibitem{nico19}
\BIBentryALTinterwordspacing
E.~Polino, M.~Riva, M.~Valeri, R.~Silvestri, G.~Corrielli, A.~Crespi,
  N.~Spagnolo, R.~Osellame, and F.~Sciarrino, ``Experimental multiphase
  estimation on a chip,'' \emph{Optica}, vol.~6, no.~3, pp. 288--295, Mar 2019.
  [Online]. Available:
  \url{http://www.osapublishing.org/optica/abstract.cfm?URI=optica-6-3-288}
\BIBentrySTDinterwordspacing

\bibitem{animesh}
\BIBentryALTinterwordspacing
F.~Albarelli, J.~F. Friel, and A.~Datta, ``{Evaluating the Holevo Cramér-Rao
  bound for multi-parameter quantum metrology},'' arXiv:1906.05724. [Online].
  Available: \url{https://arxiv.org/abs/1906.05724}
\BIBentrySTDinterwordspacing

\bibitem{paris01}
\BIBentryALTinterwordspacing
M.~G.~A. Paris, G.~M. D'Ariano, and M.~F. Sacchi, ``Maximum-likelihood method
  in quantum estimation,'' \emph{AIP Conference Proceedings}, vol. 568, no.~1,
  pp. 456--467, 2001. [Online]. Available:
  \url{https://aip.scitation.org/doi/abs/10.1063/1.1381908}
\BIBentrySTDinterwordspacing

\bibitem{john18}
\BIBentryALTinterwordspacing
J.~M. Donohue, V.~Ansari, J.~\ifmmode \check{R}\else
  \v{R}\fi{}eh\'a\ifmmode~\check{c}\else \v{c}\fi{}ek, Z.~Hradil, B.~Stoklasa,
  M.~Pa\'ur, L.~L. S\'anchez-Soto, and C.~Silberhorn, ``Quantum-limited
  time-frequency estimation through mode-selective photon measurement,''
  \emph{Phys. Rev. Lett.}, vol. 121, p. 090501, Aug 2018. [Online]. Available:
  \url{https://link.aps.org/doi/10.1103/PhysRevLett.121.090501}
\BIBentrySTDinterwordspacing

\bibitem{qcrbxstine}
\BIBentryALTinterwordspacing
M.~Tsang, R.~Nair, and X.-M. Lu, ``Quantum theory of superresolution for two
  incoherent optical point sources,'' \emph{Phys. Rev. X}, vol.~6, p. 031033,
  Aug 2016. [Online]. Available:
  \url{https://link.aps.org/doi/10.1103/PhysRevX.6.031033}
\BIBentrySTDinterwordspacing

\bibitem{luisxstine2}
\BIBentryALTinterwordspacing
J.~Rehacek, M.~Pa\'{u}r, B.~Stoklasa, Z.~Hradil, and L.~L. S\'{a}nchez-Soto,
  ``Optimal measurements for resolution beyond the rayleigh limit,'' \emph{Opt.
  Lett.}, vol.~42, no.~2, pp. 231--234, Jan 2017. [Online]. Available:
  \url{http://ol.osa.org/abstract.cfm?URI=ol-42-2-231}
\BIBentrySTDinterwordspacing

\bibitem{luisxstine}
\BIBentryALTinterwordspacing
J.~\ifmmode \check{R}\else \v{R}\fi{}eha\ifmmode~\check{c}\else \v{c}\fi{}ek,
  Z.~Hradil, B.~Stoklasa, M.~Pa\'ur, J.~Grover, A.~Krzic, and L.~L.
  S\'anchez-Soto, ``Multiparameter quantum metrology of incoherent point
  sources: Towards realistic superresolution,'' \emph{Phys. Rev. A}, vol.~96,
  p. 062107, Dec 2017. [Online]. Available:
  \url{https://link.aps.org/doi/10.1103/PhysRevA.96.062107}
\BIBentrySTDinterwordspacing

\bibitem{eckstein11}
\BIBentryALTinterwordspacing
A.~Eckstein, B.~Brecht, and C.~Silberhorn, ``A quantum pulse gate based on
  spectrally engineered sum frequency generation,'' \emph{Opt. Express},
  vol.~19, no.~15, pp. 13\,770--13\,778, Jul 2011. [Online]. Available:
  \url{http://www.opticsexpress.org/abstract.cfm?URI=oe-19-15-13770}
\BIBentrySTDinterwordspacing

\bibitem{rubio19}
\BIBentryALTinterwordspacing
J.~Rubio and J.~Dunningham, ``Quantum metrology in the presence of limited
  data,'' \emph{New Journal of Physics}, vol.~21, no.~4, p. 043037, apr 2019.
  [Online]. Available: \url{https://doi.org/10.1088%2F1367-2630%2Fab098b}
\BIBentrySTDinterwordspacing

\bibitem{lumino18}
\BIBentryALTinterwordspacing
A.~Lumino, E.~Polino, A.~S. Rab, G.~Milani, N.~Spagnolo, N.~Wiebe, and
  F.~Sciarrino, ``Experimental phase estimation enhanced by machine learning,''
  \emph{Phys. Rev. Applied}, vol.~10, p. 044033, Oct 2018. [Online]. Available:
  \url{https://link.aps.org/doi/10.1103/PhysRevApplied.10.044033}
\BIBentrySTDinterwordspacing

\bibitem{wiebe16}
\BIBentryALTinterwordspacing
N.~Wiebe and C.~Granade, ``Efficient bayesian phase estimation,'' \emph{Phys.
  Rev. Lett.}, vol. 117, p. 010503, Jun 2016. [Online]. Available:
  \url{https://link.aps.org/doi/10.1103/PhysRevLett.117.010503}
\BIBentrySTDinterwordspacing

\bibitem{pso1}
\BIBentryALTinterwordspacing
A.~Hentschel and B.~C. Sanders, ``Machine learning for precise quantum
  measurement,'' \emph{Phys. Rev. Lett.}, vol. 104, p. 063603, Feb 2010.
  [Online]. Available:
  \url{https://link.aps.org/doi/10.1103/PhysRevLett.104.063603}
\BIBentrySTDinterwordspacing

\bibitem{pso2}
\BIBentryALTinterwordspacing
------, ``Efficient algorithm for optimizing adaptive quantum metrology
  processes,'' \emph{Phys. Rev. Lett.}, vol. 107, p. 233601, Nov 2011.
  [Online]. Available:
  \url{https://link.aps.org/doi/10.1103/PhysRevLett.107.233601}
\BIBentrySTDinterwordspacing

\end{thebibliography}

\end{document}